\documentclass[twocolumn]{aastex631}

\shorttitle{Conformity Across Cosmic Time}
\shortauthors{Olsen \& Gawiser}
\graphicspath{{./}{figures/}}

\begin{document}

\title{Searching for Conformity Across Cosmic Time with Local Group and Local Volume Star Formation Histories}

 \correspondingauthor{Charlotte Olsen}
\email{olsen@physics.rutgers.edu}

\author[0000-0002-8085-7578]{Charlotte Olsen}
\affiliation{ 
Department of Physics and Astronomy, Rutgers University,  
Piscataway, NJ 08854, USA}

\author[0000-0003-1530-8713]{Eric Gawiser}
\affiliation{ 
Department of Physics and Astronomy, Rutgers University,  
Piscataway, NJ 08854, USA}

\begin{abstract}
Conformity denotes the correlation of properties between pairs of galaxies as a function of separation. Correlations between properties such as star formation rate (SFR), stellar mass, and specific star formation rate (sSFR) have implications for the impact of environment upon galaxy formation and evolution. Conformity between primary galaxies and satellites within the same dark matter halo has been well documented in simulations and observations. However, the existence of conformity at greater distances -- known as two-halo conformity -- remains uncertain. We investigate whether galaxies in the Local Volume to a distance of 4 Mpc show conformity by examining SFR, sSFR, stellar mass, and quenched fraction as a function of physical separation. Making use of the star formation histories of these galaxies, we then extend this analysis back in time to offer the first probe of conformity inside our past light cone. At the present day, we find that the stellar mass or sSFR of a galaxy correlates with the median SFR of neighboring galaxies at a separation of 2 to 3 Mpc.  At a lookback time of 1 Gyr, we find a correlation with the quenched fraction of neighboring galaxies, again at 2 to 3 Mpc separation. These signals of conformity likely arise from the differences between the recent star formation histories of Local Group dwarf galaxies and those outside the Local Group. As current and future missions including JWST, Rubin, and Roman expand the sample of Local Volume galaxies, tests of conformity using star formation histories will provide an important tool for exploring spatio-temporal correlations between galaxies. 

\end{abstract}


\section{Introduction} \label{sec:intro}
The Milky Way (MW) and its satellites offer a valuable laboratory for studying galaxies. The ability to resolve individual stars and stellar clusters reveals detailed insights into how galaxies in the Local Group (LG) have built up stellar mass over their lifetimes. While this picture within the Milky Way system may be incomplete due to dust obscuration and our position within the Galactic disk, studies of the Milky Way have driven how we model evolution in Milky Way like galaxies and their satellites. Indeed, many simulations model Milky Way like systems to compare against observations of the Milky Way and Andromeda (M31) in order to better understand the evolution of galaxies and their satellites \citep[e.g., ][]{Benson2002, Brooks2013, Garrison-Kimmel2014, Deason2015, Hopkins2014, Hopkins2018, Applebaum2021, Grand2021}. Of particular interest is what the satellites of these systems can tell us about how the interplay between the primary galaxy and its satellites influences galaxy evolution. Recent efforts to build a statistically significant sample of Milky Way like galaxies and their satellites offer exciting prospects for studying the environmental drivers of evolution of Milky Way like galaxy systems \citep{Geha2017,Mao2021,Carlsten2022}.  Large surveys such as the Sloan Digital Sky Survey (SDSS: \cite{SDSSYork}), CANDELS \citep{Grogin2011CANDELS,Koekemoer2011CANDELS}, the forthcoming Rubin-LSST \citep{Ivezic2019}, etc. offer a similar opportunity to explore environmental effects on the evolution of various types of galaxies and their satellites. 

One way in which large surveys and simulations have been used to probe the role of environment in galaxy evolution is by studying in situ properties as a function of separation.The correlation between observable physical properties of a primary galaxy and its neighbors is `galactic conformity' as first defined by \cite{Weinmann2006}. These properties include star formation rate (SFR), color, specific star formation rate (sSFR), gas mass, stellar mass, and quenched fraction. Tests for conformity have shown such trends as neighbors of quenched, passive, or red galaxies being more likely to be quenched, passive, or red themselves \citep{WangWhite2012, Phillips2014, Knobel2015, Hartley2015, Kawinwanichakij2016, Ayromlou2022}. \cite{Kauffmann2013} found that low mass galaxies are more likely to have low sSFR neighbors, and that this effect extends beyond the dark mater halo of the galaxy up to a separation of 4 Mpc. Although controversial, additional studies have presented evidence for galactic conformity between primaries and galaxies outside their halos \citep{Kauffmann2015, Campbell2015, HearinWatson2015, Bray2016, Lacerna2018, Lacerna2022, Ayromlou2022}. This is often referred to as `two-halo conformity' where a correlation within the dark matter halo is `one-halo conformity.' In these studies, one-halo conformity is generally a much stronger signal, and drops off significantly in the two-halo regime. Disagreements between models, simulations, and observations could be indicative of models lacking realistic physical processes \citep{Kauffmann2015}, or could indicate sensitivity to selection criteria in observations or projection effects \citep{Bray2016}. Studies reproducing the results of \cite{Kauffmann2013} have shown that much of their two-halo signal is a result of contamination of the isolation criterion\citep{Tinker2017,Sin2017}.  
 
While the Local Group is limited in the number of galaxies available for study, it has advantages over larger surveys in resolution and sensitivity. Nearby galaxies can be studied in detail through resolved imaging, and faint field galaxies which would be below the detection limit in current large surveys can be observed, allowing for more types of environments to be probed. The Local Group (D$\leq 1$ Mpc) hosts mainly M31, the Milky Way and their satellites and is at the center of the Local Volume (LV). The LV offers a variety of environments to study, from field galaxies to groups.

While the Local Volume offers a large selection of groups, only the volume within $\sim 4$ Mpc is close enough to resolve individual stars thereby creating Color-Magnitude Diagrams \citep[CMDs: e.g.][]{dalcanton}. The ACS Nearby Galaxy Survey Treasury (ANGST, \citealt{dalcanton}) observations expand upon those available for the Local Group, adding resolved observations for an additional 70 galaxies outside the Local Group. The ANGST sample contains a diverse set of galaxies that, while nearby enough to be resolved, are also distant enough for integrated photometric observations.  \cite{Olsen2021} used these observations to compare the SFHs of Local Volume dwarf galaxies out to $\sim$4 Mpc from two SFH reconstruction methods corresponding to the two types of observations. The first method is the CMD reconstruction method of \cite{dolphin} using HST images from the ANGST survey as shown in \cite{weisz}, and the second using the \texttt{Dense Basis} nonparametric Spectral Energy Distribution (SED) fitting code \citep{Iyer2019} that fit matched photometry from \cite{johnson} for 36 of the galaxies with good wavelength coverage. They found overall good agreement between the two techniques, which have complementary systematics. When studying the overall shape of the SFHs for the population of dwarf galaxies by taking the median of the normalized SFHs of all galaxies from both methods, they found synchronized star formation within the galaxies. The SFHs of these Local Volume dwarfs showed a decrease of star formation at roughly 6 Gyr lookback time followed by an increase at 3 Gyr lookback time, the cause of which is still unknown.

To further explore how star formation histories may reveal a relationship between galaxies in the Local Volume, we adopt methods for testing for conformity to the sample of the Local Volume out to 4 Mpc, and make the first conformity analysis extending back in time. Where previous tests of conformity measured the properties of primaries and neighbors (termed ``secondaries'')  as a function of separation only at the time of observation, the usage of SFHs enables us to determine these properties at past epochs. We choose times that sample the epochs of interest in  \cite{Olsen2021}, using 1 Gyr, 2 Gyr, and 4 Gyr lookback times as well as the time of observation. This analysis unifies the use of SFHs as a probe of a galaxy's evolution over time with the dependence of a galaxy's evolution to its spatial relation to other galaxies revealed by conformity tests. These two techniques in combination offer a spatio-temporal probe of galaxy evolution. Using these we will test to see if the synchronized star formation seen in the SFHs of ANGST galaxies correspond to either one or two-halo conformity.  

In Section~\ref{sec:data} we introduce the data, while Section~\ref{sec:methods} details our methodology.  In Section~\ref{sec:stellar_mass}, we discuss our results for the stellar mass of secondaries when splitting primary galaxies by stellar mass and sSFR. In Sections~\ref{sec:star_formation_rate}-
\ref{sec:qf} we offer the same analysis for SFR, sSFR, and quenched fraction of secondaries, respectively.  We discuss the implications of our results in Section~\ref{sec:discussion}, including comparing the contributions of the Local Group and Local Volume in Section~\ref{sec:LGLV},
and Section~\ref{sec:conclusions} concludes.

\section{Data}\label{sec:data}
To enable conformity tests within a contiguous volume, we combine the 36 Local Volume $<$4 Mpc ANGST galaxies from \cite{Olsen2021} with Local Group galaxies, bridging the first Mpc gap in the volume. We selected galaxies in the Local Group for which there are public SFHs determined via CMDs, and from those galaxies we included only galaxies at least as massive as the lowest mass dwarf galaxy in our Local Volume sample.  Without this mass cut,  our Local Group sample would have been  dominated by galaxies too faint to have been studied in the Local Volume. Adding 17 Local Group galaxies meeting this criterion yields a total sample of 53 galaxies.
\begin{figure}
    \centering  \includegraphics[width=.55\textwidth]{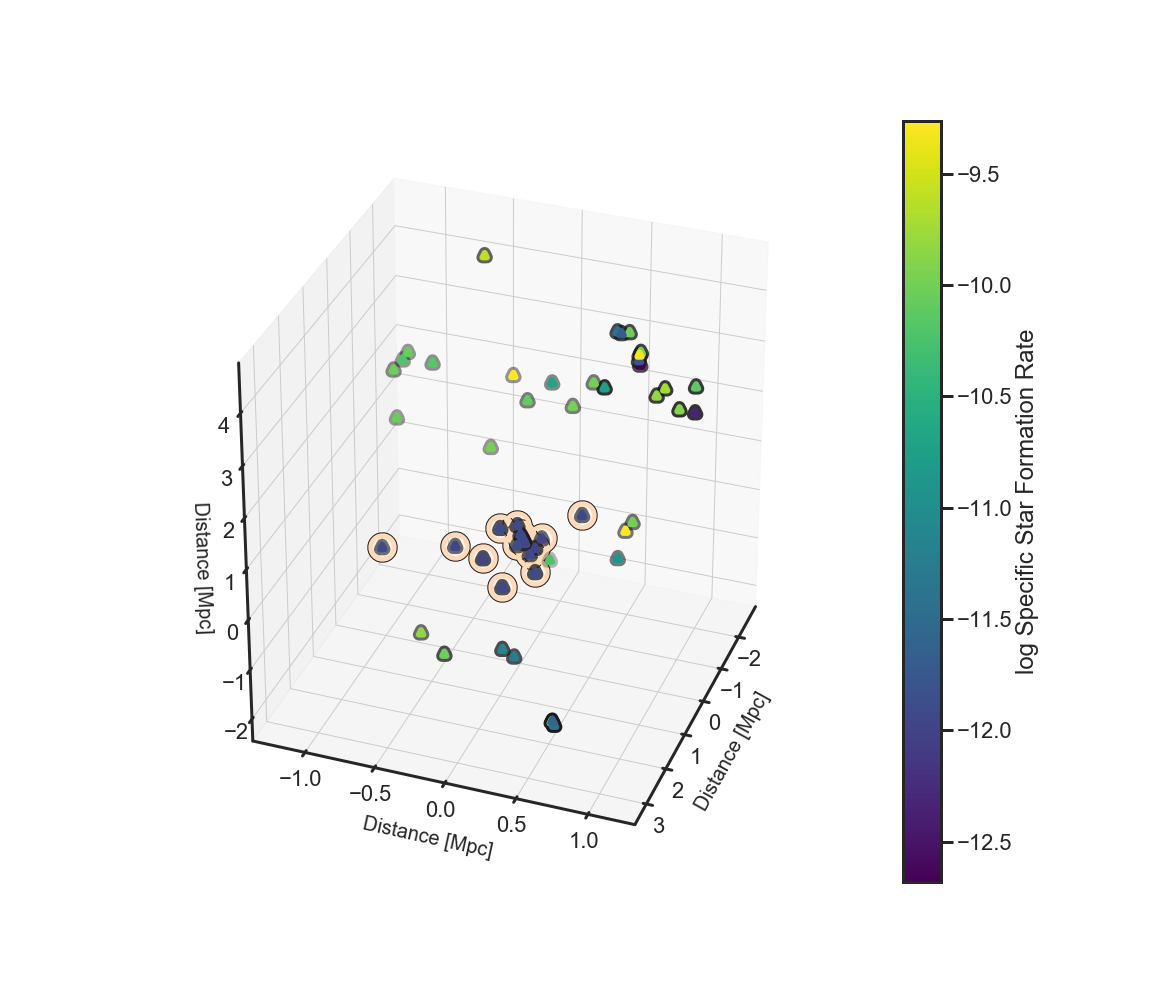}
    \caption{3D depiction of the sample colored by log$_{10}$ log$_{10}$  specific star formation rate. Local Group galaxies are shown outlined in the center of the volume. Even though the Local Group galaxies shown are ones included after a mass cut, the Local Group galaxies in the center have consistently lower sSFR than the ANGST galaxies.}
    \label{fig:3d_3pan}
\end{figure}

In Figure~\ref{fig:3d_3pan} we show  the sample colored sSFR.   
Local Group and Local Volume dwarf galaxy CMD SFHs come from \cite{WeiszlgI2014} and \cite{weisz}, respectively. While \cite{Olsen2021} found good agreement between CMD and SED derived SFHs for Local Volume galaxies, only CMD SFHs are available for Local Group galaxies. To avoid introducing biases due to methodological differences the CMD SFHs were used for all galaxies in this analysis. 
 SFHs from CMDs are generally reconstructed using bins of constant time in log space. \cite{McQuinn2010} found that starbursts were not always contained within a single bin of logarithmic width  0.1 dex. To fix this issue, they adopted a binning scheme that had bin widths of 0.3 dex in up to 5 Myr lookback time as well as from 3 to 5 Gyr lookback time. In an effort not to over-interpret the precision of the SFHs at early times while preserving the ability for the CMD to capture the stochasticity of recent star formation, the temporal binning in \cite{McQuinn2010} was adopted for the entire sample.

 Local Group SFHs are taken directly from \cite{WeiszlgI2014}, where SFHs are reported as cumulative mass formed with a precision of 2 decimal places for reported mass values and uncertainties. Occasionally multiple SFHs were reported for a single galaxy if CMDs were taken for multiple fields. In these cases the SFHs were summed.  In cases where the footprint failed to cover the whole galaxy, the total (stellar) mass was taken from the literature values measured from the integrated light, and each SFH was renormalized to produce this observed mass. The median factor by which a galaxy was renormalized is 3.7, but a small handful of galaxies were renormalised by a factor of over ten. In all cases we made the typical assumption that the available SFH is representative of the whole galaxy with the caveat that this may not be the case, particularly when the footprint of the CMD covers a small area. While uncertainties were reported for the publicly available SFHs for the (cumulative) stellar mass formed  within each time bin, the lack of reported covariances means that we cannot account for correlations between bins when computing the differential SFHs used in this analysis.   
  Where needed, binned values of stellar mass formed were combined into larger time bins, with an unweighted sum of variances used to model the uncertainty on these combined quantities.

\section{Methodology}\label{sec:methods}

\subsection{Tests of Conformity}
We will test for conformity using the method from \cite{Kauffmann2013}.  Their isolation criteria for a galaxy to be considered a primary was that there be no galaxies more massive than half the mass of the selected galaxy within a projected separation of 500 kpc and a velocity interval of $\Delta$v $=$ 500 km/s. We use available literature values of distances to our galaxies to modify \cite{Kauffmann2013}'s selection criterion to be based on physical distance and not projected distance and avoid possible contamination. This allows us to adopt an isolation criterion such that each galaxy is a primary as long as there are no galaxies within 500 kpc of \textit{physical} distance of the primary with more than half its stellar mass. Many studies using large surveys successfully utilize projected distances with careful selection criteria, but the small size of our sample further motivates our use of physical distances as our number of primaries drops from 18 to 13 when switching to a criterion using projected distances. Additionally, some of our galaxies are members of groups. Massive centrals not included in the sample of available SFHs are included in our set of primaries to ensure that their lower mass satellites do not contaminate the primary sample.

\begin{figure}
    \centering \includegraphics[width=0.46\textwidth]{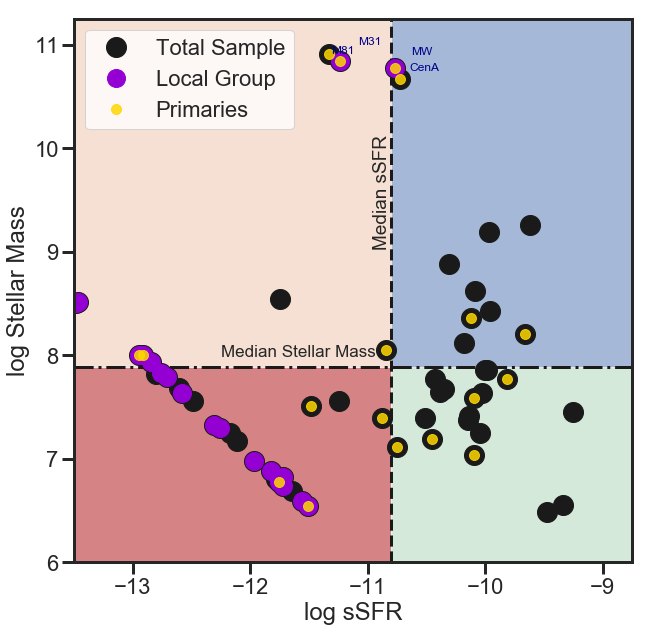}
    \caption{The distribution of primaries and secondaries as a function of stellar mass and sSFR at the time of observation. We see very little correlation between stellar mass and sSFR except for Local Group dwarfs, which typically have low SFR at the time of observation.   
    Primaries chosen by our isolation criterion are shown in yellow. The median stellar mass and median sSFR are shown in the dot-dashed and dashed lines to show where primary galaxies reside above and below 50th percentile cuts.Colors are to guide the eye and delineate what quadrants of galaxies occupy on stellar mass/sSFR space.}
    \label{fig:stellar_mass_sSFR}
\end{figure}

To determine whether the galaxy property of a secondary galaxy is correlated with its primary, we split our sample of primary galaxies on stellar mass and sSFR. In Figure~\ref{fig:stellar_mass_sSFR} we show the stellar mass and sSFR of all galaxies in our sample. We see no correlation between  the stellar mass and sSFR of our galaxies except in the case where there has been little very recent star formation and an upper limit was used for SFR, as is seen in the lower left quadrant of the figure.  

\begin{figure}
    \centering    \includegraphics[width=0.46\textwidth]{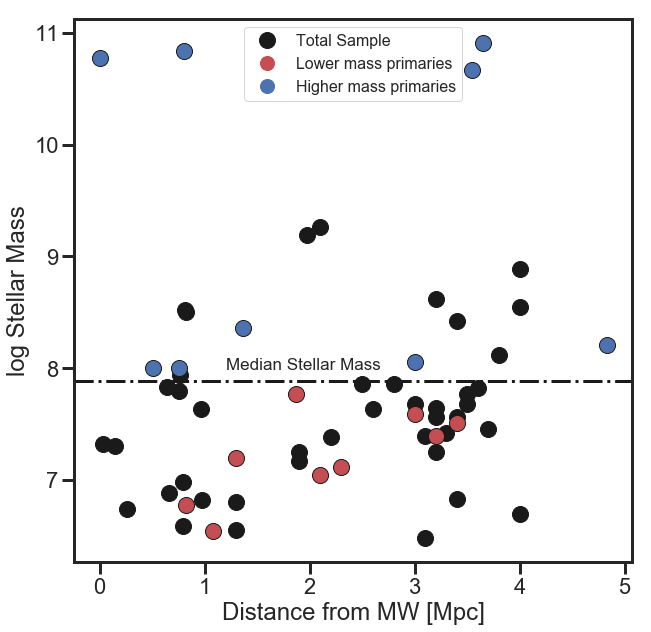}
    \caption{Log stellar mass of galaxies as a function of distance from the Milky Way. Secondaries are shaded in black, while the upper 50th percentile of stellar mass primaries are shown in blue and the lower in red.}
    \label{fig:Distance_mstar}
\end{figure}

In Figure~\ref{fig:Distance_mstar} we look at the log stellar mass of all galaxies as a function of distance from the Milky Way. We color the higher stellar mass primaries blue, and lower stellar mass primaries red. We see no strong radial trends in stellar mass. 

For each primary galaxy we define all other galaxies as secondaries and bin all secondaries by separation in Mpc from the primary. Distance bin edges were chosen ahead of time to be round numbers, with entire 1-halo regime in first bin, and were not varied after analyzing the data. The density of galaxy pairs in bins of increasing separation are 104, 161, 209, and 465. To best explore the relationship between the primary dark matter halo and neighboring halos, we could further require all secondaries to be of lower stellar mass than the primary, but if we try this we find many of the separation bins to be too sparsely populated to yield robust statistics. This is likely due to the bulk of our sample of primaries being low mass galaxies. The quenched fraction is calculated for all secondaries of a given set of primaries within a given distance bin.  Quenching is often determined by a cut on sSFR such that galaxies lying below that threshold are considered quenched, but we used the definition from \cite{Pacifici2016a} that a galaxy is quenched when sSFR$<0.2/t_U(z)$ where $t_U(z)$ is the age of the universe at redshift $z$. This accounts for the redshift evolution of quenched galaxies and allows us to calculate the quenched fraction of our sample at any step backwards in time. At small lookback times this is close to the sSFR = 1e-11 threshold typically used at the time of observation, but progresses modestly with each timestep. At 4 Gyr lookback time the threshold is sSFR =2e-11. Uncertainties in the quenched fraction were calculated based on Poisson uncertainties in the number of quenched secondaries in each bin corresponding to each split of primaries. 

We determine the median value of the observed secondary property in each radial bin of separation; this is a more robust statistic than the mean, especially in circumstances where a large fraction of the galaxies have upper limits on their star formation rates. Uncertainties on the binned medians are calculated using the standard deviation of the median, 
\begin{equation}
\sigma_{ \bar{x} } = \sqrt {\frac{ < \sigma^2_{x_i}>}{n}}
\sqrt{\frac{\pi (2n+1)}{4n}} 
\end{equation} 
where $<\sigma^2_{x_i}>$ is the average reported variance of the property across individual galaxies and $n$ is the total number of galaxies; the first term gives the standard deviation of the mean for heteroscedastic uncertainties, and the second term corrects for the reduced efficiency of the median. 

Conformity is indicated in a particular bin of separation if the median property of secondaries of one set of primaries (e.g., high stellar mass) shows a statistically significant difference from the median property of secondaries of the complementary set of primaries.

\subsection{Utilizing the Star Formation History}
Tests of conformity have traditionally been done at the time of observation, but SFHs allow us to test for conformity at multiple epochs.  We include 4 time steps. The first is the time of observation, which represents the `traditional' time step for conformity tests. The next time step probes conformity at 1 Gyr lookback time. Our third time step is at 2 Gyrs lookback time, with the final time step at 4 Gyrs lookback time. The time bins of 1, 2, and 4, Gyr lookback times were chosen a priori to investigate times at which the normalized median SFHs reported by \cite{Olsen2021} were on  sharply increasing or near minimum.

\section{Results}
 Our results center on tests of conformity for the properties of stellar mass, SFR, sSFR, and quenched fraction. We perform tests of conformity on the entire sample at 4 separate time steps and split the pairs by the stellar mass or sSFR of the primary. We split primaries on the 50th percentile as this was the most robust choice for a small sample. The signal is sensitive to the choice of percentile we pick, but only to the degree that distance bins in one of the splits become sparsely populated. For example, when splitting on the 25th percentile in primary stellar mass, the 0-25th percentile bins are sparsely populated with secondaries. In the case of splitting on stellar mass we have only 11 galaxy pairs in the first separation bin. Pairs are binned by distance of the secondary from the primary with 4 radial bins of $R\leq 1$ Mpc, $1<R\leq 2$ Mpc, $2<R\leq 3$ Mpc, and $R>3$ Mpc. The maximum separation between pairs is $6.6$ Mpc. A separation in the first bin of $R\leq 1$ Mpc probes one-halo conformity, while separation in other bins falls into the two-halo conformity regime. As we look back in time, these separations represent comoving distances based on the galaxies' present-day positions. We do not attempt to account for migration of galaxies between radial bins;this may cause the separation of some $<1$ Mpc pairs to be underestimated if satellites had not yet accreted as an earlier epoch. That could cause two-halo conformity to alias to one-halo conformity, but not vice versa. 
 Figures ~\ref{fig:mstar_mstar}, ~\ref{fig:sfr_sfr}, ~\ref{fig:ssfr_ssfr}, and ~\ref{fig:qfrac_qfrac} show 128 tests of conformity in total. If  our uncertainties were correctly estimated and gaussian distributed, we would expect 6 spurious detections at the 2$\sigma$ level and 1 at the 2.5$\sigma$ level. However, since our uncertainties appear to be significantly overestimated (see Figure ~\ref{fig:sig_hist} and accompanying discussion), we define a probable (possible) detection of conformity as a $>$2.5$\sigma$ ($>$2.0$\sigma$) difference between properties of the secondaries of low stellar mass/sSFR galaxies vs those of high stellar mass/sSFR galaxies.

\subsection{Search for Conformity Split on Stellar Mass}\label{sec:stellar_mass}
We take all galaxies defined by our isolation criterion as primaries and divide these into two groups that lie below and above the median stellar mass. Distances from every primary are then calculated to every other galaxy. For all pairs the stellar mass of these secondary galaxies is binned by distance from the primary. This same procedure is then carried out where primaries are instead split on sSFR instead of mass.

\begin{figure*}
    \centering
    \includegraphics[width=0.95\textwidth]{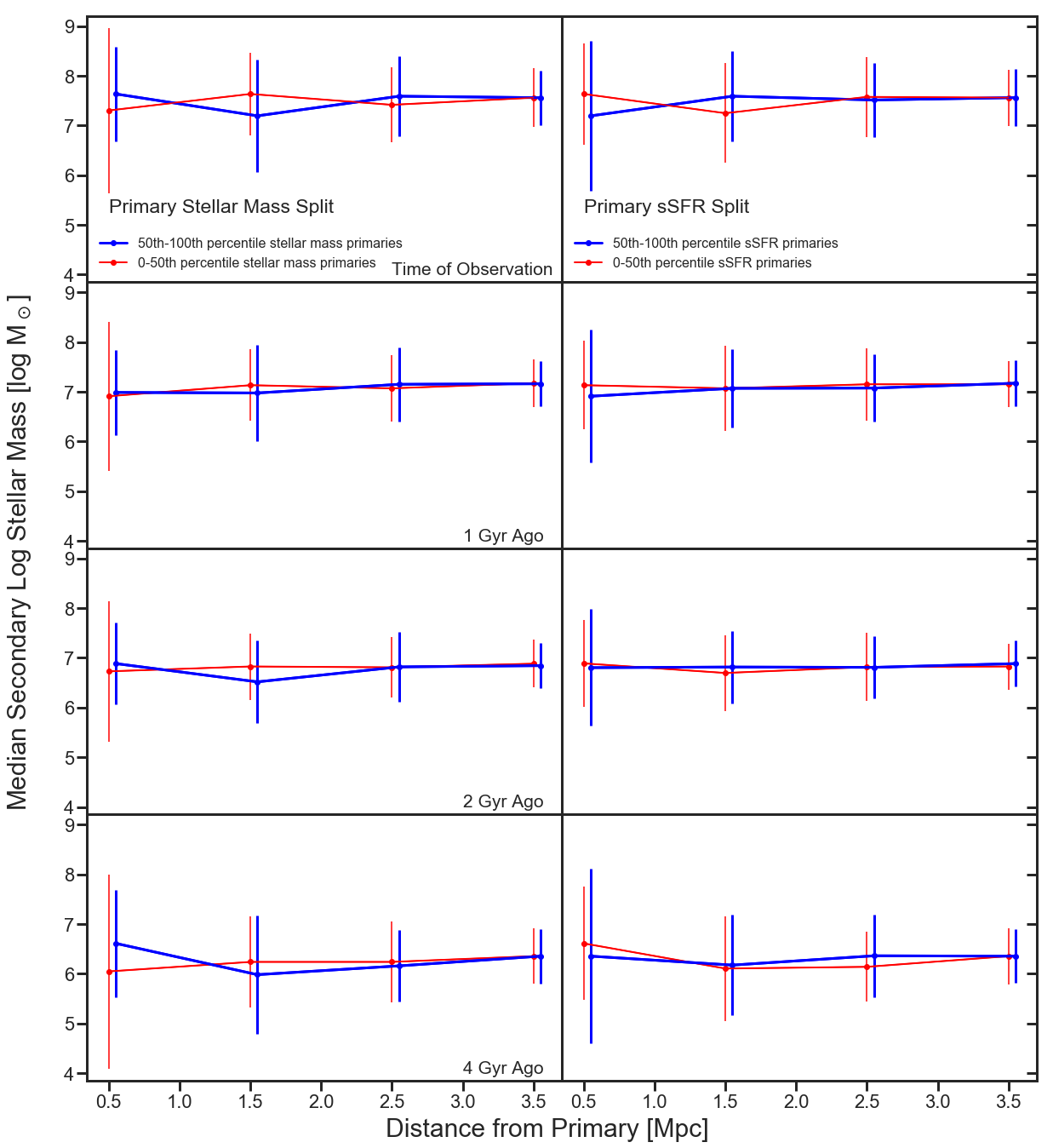}
    \caption{Blue and red depict galaxy pairs whose primaries are split into two subsets representing those greater than and less than the median stellar mass (left) or sSFR (right) of all primaries at the time of observation. Both panels show the median stellar mass for  secondary galaxies of these subsets as a function of radial separation from the primary, with the top panel the relation at the time of observation, and the subsequent panels showing lookback times of 1, 2, and 4 Gyrs. The stellar mass of secondaries does not differ significantly based on the stellar mass or sSFR of their primaries at any separation or lookback time. 
    }
    \label{fig:mstar_mstar}
\end{figure*}

In Figure ~\ref{fig:mstar_mstar} we compare the 
 median stellar mass of secondaries based on the separation from the primary. On the left we show the sample split by the stellar mass of the primary, and on the right we show the sample split by the sSFR of primaries. For both columns we see a steady increase in stellar mass with each increasing time step, but no separation between the high and low mass/sSFR splits. This indicates that the stellar mass of secondaries is not sensitive to the stellar mass and sSFR of primaries.

\subsection{Search for Conformity Split on Star Formation Rate}\label{sec:star_formation_rate}

\begin{figure*}
    \centering   \includegraphics[width=0.95\textwidth]{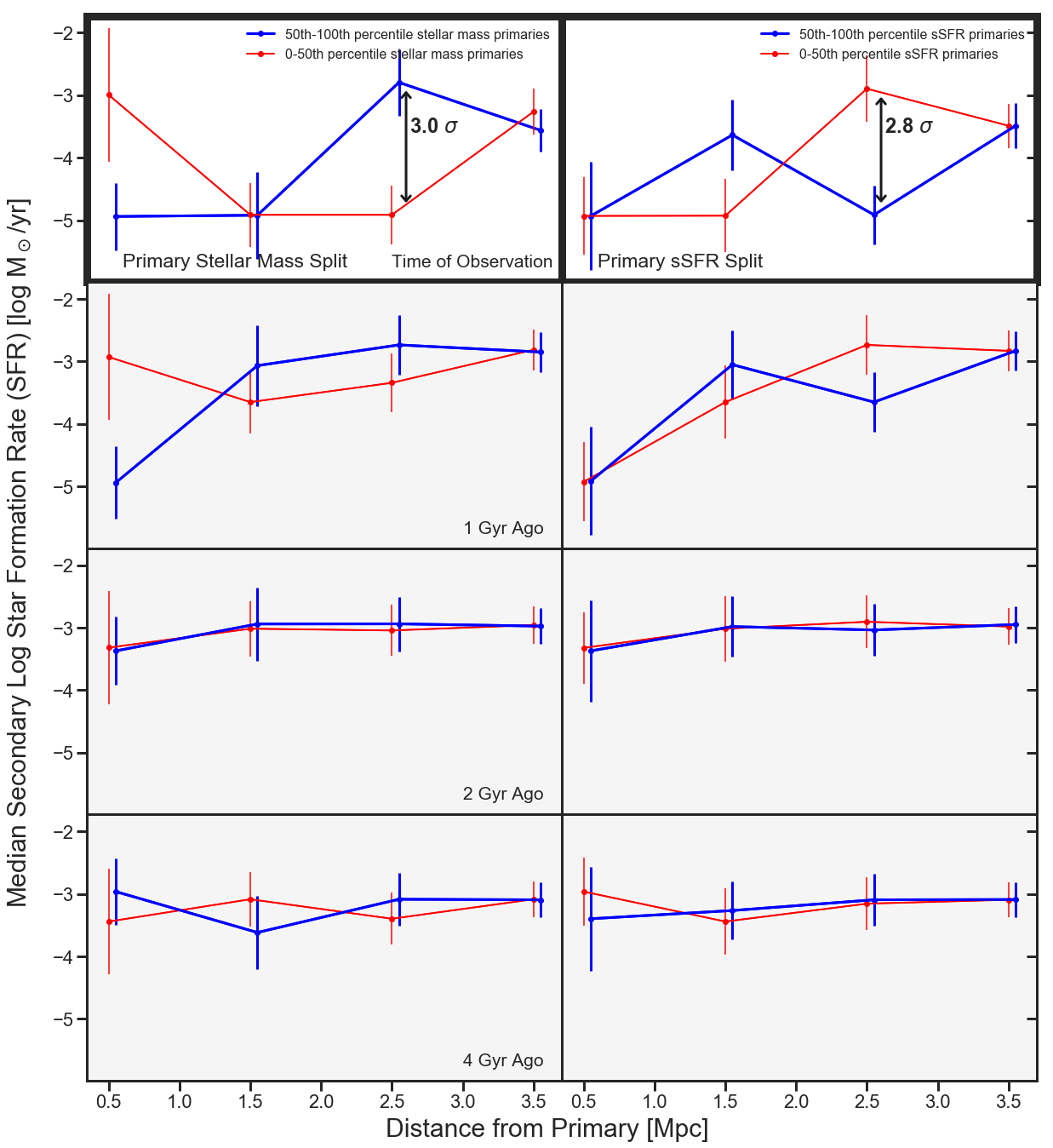}
    \caption{
    Blue and red depict galaxy pairs whose primaries split into two subsets representing those greater than and less than the median stellar mass (left) or sSFR (right) of all primaries at the time of observation. At the time of observation, secondaries of primaries split on stellar mass show  conformity of 3.0 sigma significance at separations of between 2 and 3 Mpc. At this same distance separation secondaries of low sSFR primaries have higher SFR than their counterparts to a significance of 2.8 sigma. At 1 Gyr, we still see marginally elevated SFRs for secondaries of high stellar mass and low sSFR primaries, but no other trends remain. At lookback times of 2 and 4 Gyrs there is no appreciable difference at any separation.}
    \label{fig:sfr_sfr}
\end{figure*}

We see differences between secondary SFR as a function of separation when primaries are split on median stellar mass and when primaries are split on median sSFR at the time of observation, but not at 2 and 4 Gyrs lookback time. 
In the left panel of Figure~\ref{fig:sfr_sfr} we see a signal of conformity at the time of observation, showing secondaries of high mass primaries at 2-3 Mpc separations have higher SFRs compared to secondaries of low mass primaries. When splitting on primary sSFR in the right panel of Figure~\ref{fig:sfr_sfr} we see that secondaries with lower sSFR primaries have median elevated SFRs between 2 and 3 Mpc at the time of observation with a 2.8$\sigma$ significance. Interestingly, at the time of observation, both splits show a strong crossing behavior between 1 and 2 Mpc. At 1 Gyr lookback time we see similar behavior as seen at the time of observation, but only marginally. Also of interest is that relatively high median SFRs are associated with high stellar mass and low sSFR primaries and that while there is a suggestion of it for the stellar mass split, there is not a significant detection of conformity in the one-halo regime.

\subsection{Searching for Conformity Split on Specific Star Formation Rate}
\label{sec:ssfr}

\begin{figure*}
    \centering
    \includegraphics[width=0.95\textwidth]{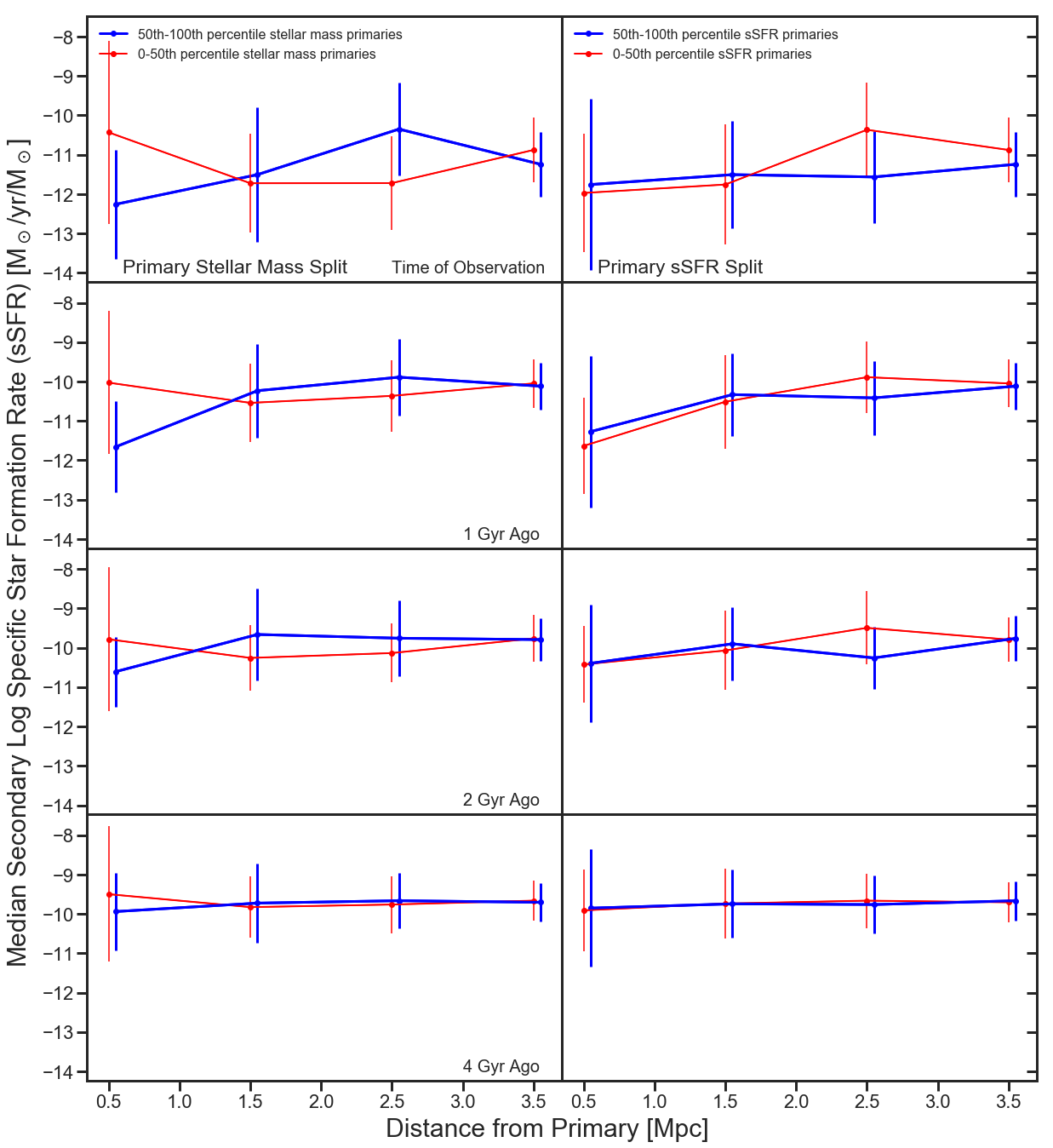}
    \caption{Blue and red depict galaxy pairs where primaries split into two subsets representing those greater than and less than the median stellar mass (left) or sSFR (right) of all primaries at the time of observation. Secondaries of primaries split on stellar mass show no appreciable difference in sSFR, though still show similar behavior to SFR with marginal separations of between 2 and 3 Mpc at the time of observation where secondaries of high stellar mass primaries have higher sSFR than their counterparts.
    Similarly, when splitting on the sSFR of the primary we see secondaries in the 2 to 3 Mpc distance bin may have marginally higher sSFR for low sSFR primaries than for high sSFR primaries. This follows the overall trends seen in the results for SFR, but only weakly. }
    \label{fig:ssfr_ssfr}
\end{figure*}
In Figure~\ref{fig:ssfr_ssfr} we show the sSFRs of secondaries in galaxy pairs split by stellar mass (left) and sSFR (right) of their primaries. The overall pattern of median sSFR in secondaries is similar to that of SFR in overall behavior, but with no significant difference. This is perhaps not surprising due to the nature of sSFR, and knowing that the stellar mass of secondaries is not sensitive to the primary properties of stellar mass and sSFR.

\subsection{Searching for Conformity Split on Quenched Fraction} \label{sec:qf}
\begin{figure*}
    \centering
    \includegraphics[width=0.95\textwidth]{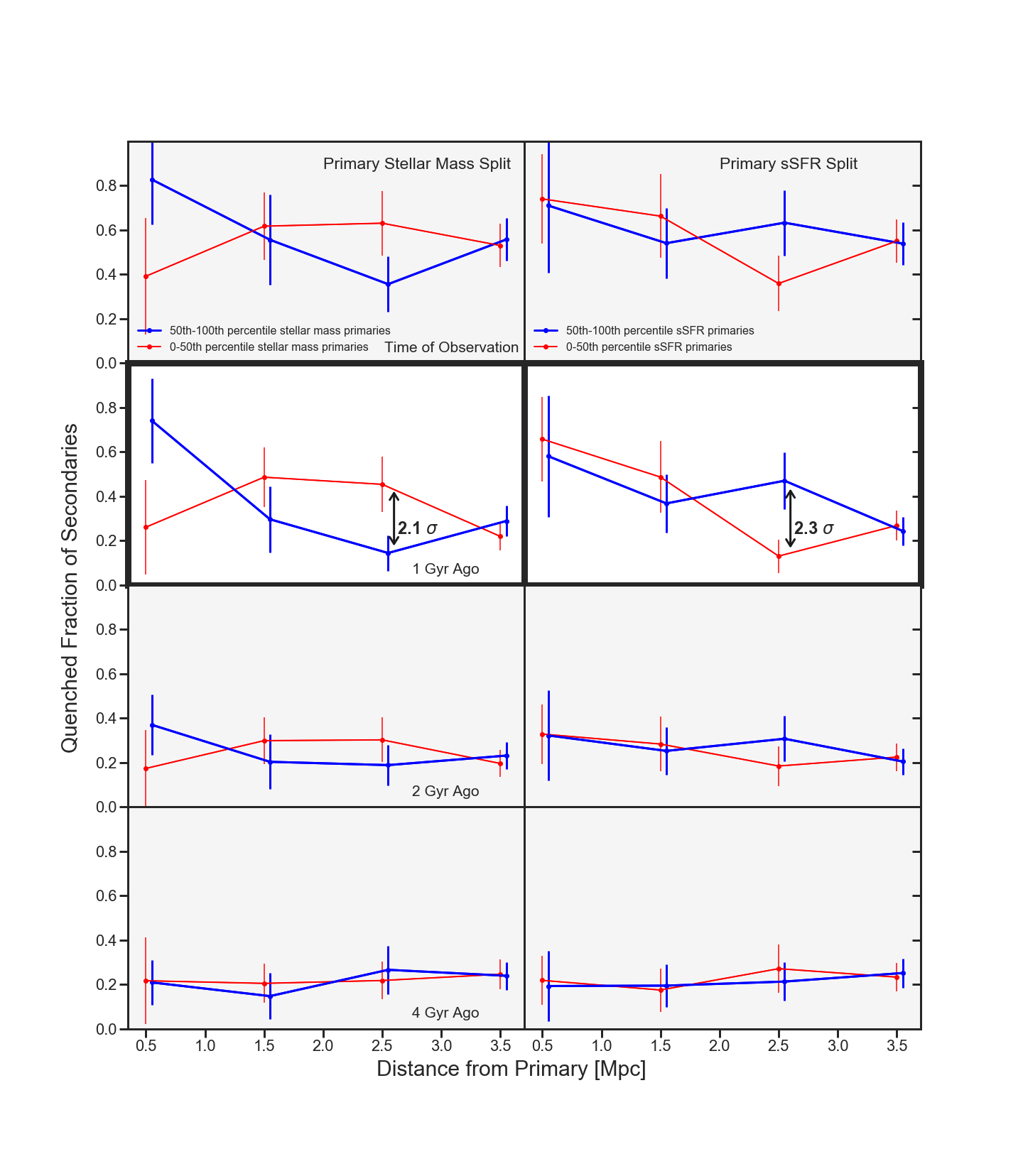}
    \caption{Blue and red depict galaxy pairs where primaries are split into two subsets representing those greater than and less than the median stellar mass (left) or sSFR (right) of all primaries at the time of observation. No conformity is seen in secondaries of primaries split on stellar mass in the oldest time steps of 2 and 4 Gyrs, but at 1 Gyr secondaries of low stellar mass primaries are likely to be quenched with a significance of 2.1 $\sigma$ when separated by 2 to 3 Mpc from the primary. This is also seen at the time of observation albeit marginally.
    The time evolution of quenched fraction of primaries split on sSFR shows similar behavior as those split on stellar mass, but with secondaries of high sSFR more likely to be quenched by a significance of 2.3 $\sigma$ when located at 2 to 3 Mpc from the primary at 1 Gyr lookback time. Again we see no conformity at 2 and 4 Gyrs lookback time.}
    \label{fig:qfrac_qfrac}
\end{figure*}

In Figure ~\ref{fig:qfrac_qfrac} we show the quenched fraction of secondaries in galaxy pairs where the primary is split by stellar mass and sSFR. In this case signals of conformity are seen at 1 Gyr lookback time with differences of 2.1 $\sigma$ and 2.3 $\sigma$ between 2 and 3 Mpc for primaries split on stellar mass and sSFR respectively. Of interest is how the higher number of quenched galaxies shows a similar switching as seen in the median SFR, but also that the correlation between high and low stellar mass and sSFR is reversed. Here we see that low stellar mass primaries have more quenched secondaries within 2 to 3 Mpc from them than their high stellar mass counterparts, as do high sSFR primaries for secondaries in the same distance bin. Again, when looking at secondaries split on primary stellar mass or sSFR, we see no sign of conformity in quenched fraction at 2 and 4 Gyrs lookback time. Seeing possible conformity in galaxy pairs in the same separation bin but different times begs the question of how much migration we expect between separation bins. Using the relative velocity of each galaxy pair to update their radial separation while keeping their transverse separation fixed, we found that  38\% of pairs in the sample would have moved enough to migrate into another bin over the course of 4 Gyr, but that number drops to 13\%  when looking at a time interval of only 1 Gyr.  Because transverse distances are also changing with time, this likely underestimates the true rate of change by a moderate amount.

\section{Discussion}
\label{sec:discussion}

\begin{figure}
    \centering    \includegraphics[width=0.48\textwidth]{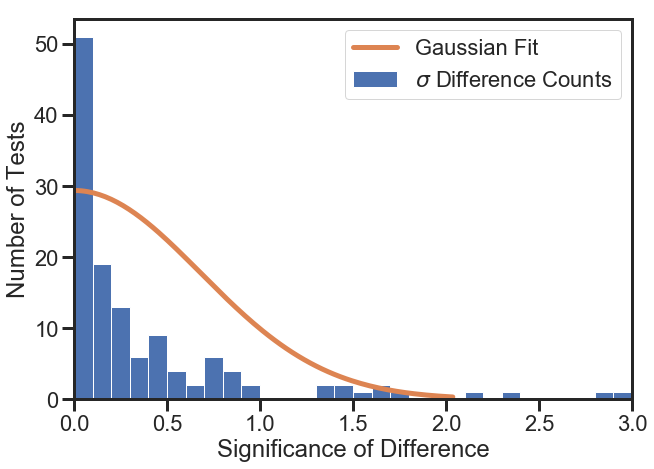}
    \caption{Frequency of significances of differences between secondary properties for all 128 tests of conformity. Our two possible detections at $\geq 2 \sigma$ as well as the two probable detections at $2.8\sigma$ and $3.0\sigma$ all appear to be robust outliers.  The Gaussian fit to the observed distribution of differences has a standard deviation of 0.5,  implying that our bin-by-bin uncertainties are roughly a factor of two larger than they should be and shows that we  expect a nominal $>2 \sigma$ difference just 0.014 times out of 128 tests, and nominal $>2.5 \sigma$ difference just 0.00017 times out of 128 tests.}
    \label{fig:sig_hist}
\end{figure}

Of 128 tests of conformity, we found two possible detections of conformity at 1 Gyr lookback time when examining the quenched fraction of secondaries separated by 2 and 3 Mpc from their primaries. We found two probable detections of conformity in the same distance bin when examining the median SFR of secondaries with a $2.8 \sigma$ and $3.0 \sigma$ significance respectively. Figure~\ref{fig:sig_hist} shows the formal significance of the difference within each bin for all 128 tests of secondary properties. 
As the 2 possible and 2 probable detections appear as strong outliers, this appears to be  a robust result. This analysis implies that the uncertainties inherited from the public SFHs might be overestimated by as much as a factor of 2. Figures~\ref{fig:mstar_mstar}-\ref{fig:qfrac_qfrac} reinforce this, with the deviations of most of the colored curves from a flat line being much smaller than would be expected for the nominal error bars.  The overestimated uncertainties likely result from starting with  cumulative SFHs and treating the  uncertainties of adjoining time bins as uncorrelated given the lack of reported covariances.  If we assume our uncertainties to be doubled, we could interpret observed uncertainties in other time steps with $>1\sigma$ as likely detections of conformity. For example, in the 2 to 3 Mpc separation bin at 1 Gyr for secondary SFR when splitting on primary sSFR, we see that the 1.35 $\sigma$ observed signal would be an indication of conformity at the $>$95\% confidence interval.

\subsection{Separation of the Local Group from the Local Volume}\label{sec:LGLV}

A detection of conformity within this volume is intriguing, and begs the question of how much the galaxies shown to have synchronized star formation in \cite{Olsen2021} may be contributing to this signal. To investigate this we decided to repeat the conformity analysis with only the Local Volume galaxies from \cite{Olsen2021} and their massive centrals to see if these galaxies were responsible for the conformity signal.

\begin{figure*}
    \centering
    \includegraphics[width = 0.95\textwidth]{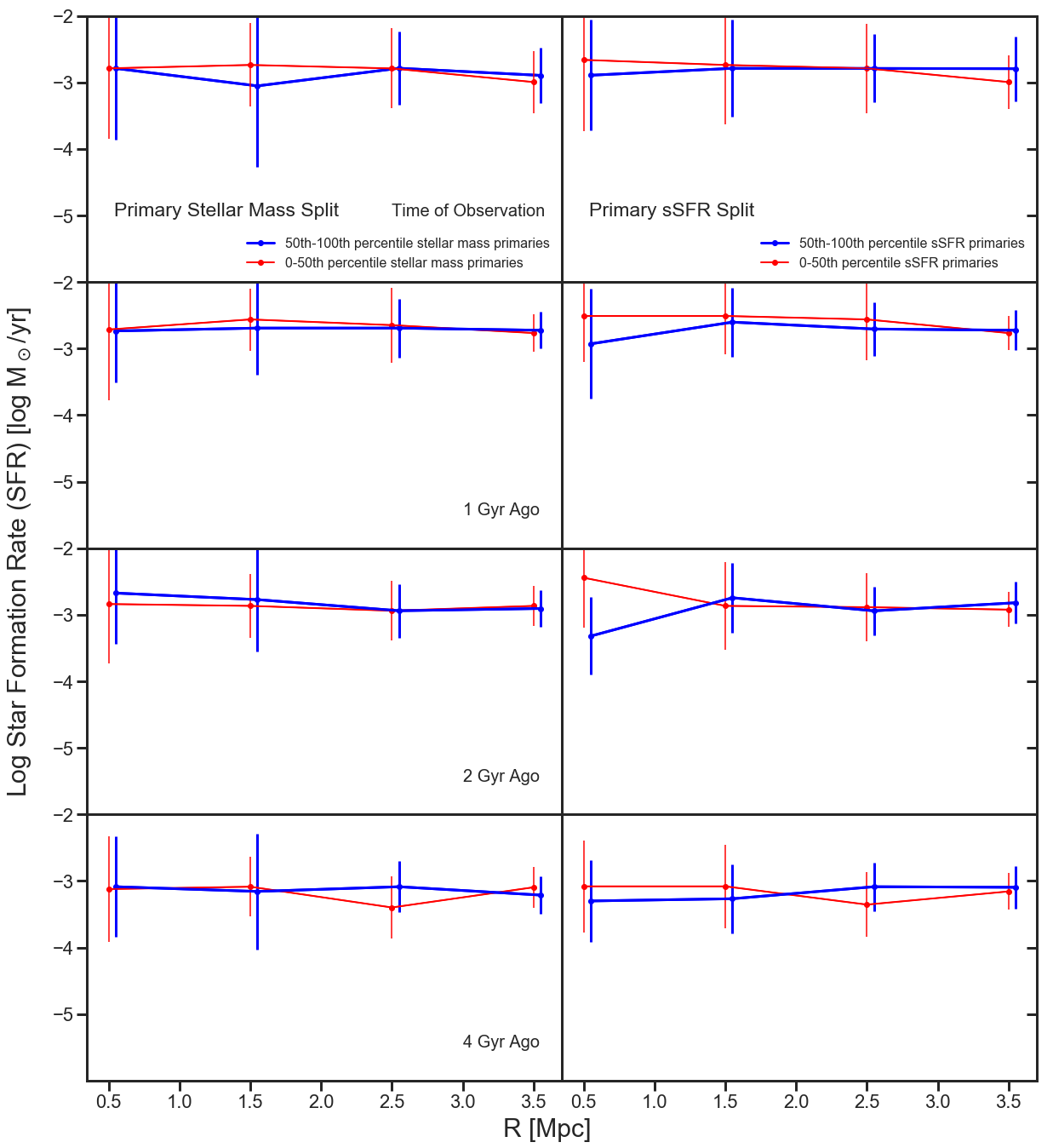}
    \caption{Removing the Local Group (LG) galaxies from our sample we plot the median SFR of secondary galaxies as a function of distance to primary galaxies. Unlike Figure ~\ref{fig:sfr_sfr} we see no separation between the median SFR of secondaries of low stellar mass or sSFR primaries of those of high stellar mass or sSFR primaries.}
    \label{fig:sfr_no_loc}
\end{figure*}
We exclude Local Group galaxies and repeat our analysis with only the ANGST galaxies from \cite{Olsen2021} and their massive centrals, with results shown  in Figure ~\ref{fig:sfr_no_loc}. Without Local Group galaxies we see no signal of conformity at the time of observation for any separation from the primary. When performing tests for secondary stellar mass, sSFR, and quenched fraction, we find similar results.

We look at similarities and differences between the ANGST and Local Group samples by qualitatively comparing the overall shape of the median of the SFHs for each sample. \cite{weisz2} compared the mean cumulative SFHs of ANGST and Local Group SFHs broken down by morphology, and found overall good agreement between the two samples when comparing the cumulative SFHs, though with some discrepancy with the dwarf irregulars where the Local Group tended to form mass on average earlier than the Local Volume dwarf irregular sample. 
\begin{figure}      
    \centering   \includegraphics[width=0.5\textwidth]{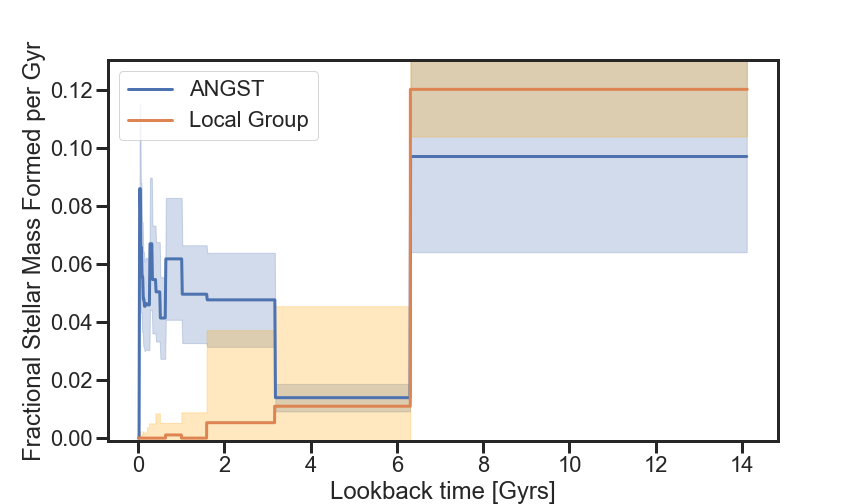}
    \caption{Median of normalized SFHs of the ANGST dwarf galaxy sample in the Local Volume, vs. the median of normalized SFHs of the Local Group dwarf galaxies. Shaded regions show 68\% scatter around the median.}
    \label{fig:LV_v_LG}
\end{figure}
Building on this work, we utilized the method from \cite{Olsen2021} to compare the overall behavior of the SFHs of the ANGST and Local Volume galaxies of \cite{weisz}. In Figure ~\ref{fig:LV_v_LG} we show the results of this method where each of the SFHs from each sample is first normalized and then the median is taken of the normalized SFHs. The 68\% scatter around the median is also shown. 

 When looking at the median of the normalized SFHs of the Local Group galaxies as compared to the Local Volume as seen in Figure~\ref{fig:LV_v_LG}, we see that the Local Group galaxies are predominantly quenched over the past Gyr. This trend, combined with the spatial location of the Local Group galaxies near the center of the Local Volume, may turn out to be the cause of conformity signals seen in our combined galaxy sample. 

 This analysis confirms the well-known literature result that,  as a whole, the Local Group galaxies formed more of their mass earlier than dwarf galaxies in the ANGST sample. Local Group galaxies are known to be especially quiescent and to have little HI  compared to the rest of the Local Volume \citep[eg, ][]{Weisz2014reion, Wetzel2015, Tollerud2018, Patel2018}. In fact, HI content in Local Group galaxies seems to drop as a function of distances to the Milky Way or M31 \citep{Putman2021}. This could be due to the interaction of the Milky Way and M31 \citep{Samuel2022}, or partially explained in the Milky Way by the interactions between smaller satellites with the Magellanic Clouds \citep[eg,][]{Jahn2022}. The Local Group dwarfs are in tension with the satellites in our sample around M83, and with satellites of galaxies found deeper in the Local Volume such as seen in the SAGA or MADCASH surveys \citep{Carlin2016, Geha2017,Mao2021, Carlin2021}. This motivates making use of simulations to see how frequently a  volume of radius 4 Mpc may be centered on a group that resembles the Local Group and generates a signal of two-halo conformity in the absence of physical efforts that cause two-halo conformity across larger volumes. If this is an infrequent phenomenon in simulations, it would argue that the Local Group-created conformity signal we have found should be studied further as a possible revelation of large-scale physical correlations in galaxy evolution.

\subsection{Implications for synchronized star formation in the Local Volume}
The lack of conformity in Figure~\ref{fig:sfr_no_loc} implies that the synchronized star formation in Local Volume dwarf galaxies is not sensitive to the properties of pairs of galaxies in any specific time step or at any distance. Since there is no significant difference between the median properties of secondaries based on the property of the primary, we can infer that synchronized star formation is not an environmental effect driven by isolated galaxies with high or low stellar mass or sSFR influencing their neighbors.

Both the Local Group and the Local Volume dwarfs show evidence of environmental effects beyond reionization, but these environmental effects in the Local Group are different from those in the Local Volume out to 4 Mpc. Further investigating the source of synchronized star formation in the Local Volume will be a challenge until more observations become available. Luckily, upcoming large scale surveys promise to fill in the volume of nearby dwarf galaxies out to greater distances and place better constraints on cosmology. Another promising area of inquiry is hydrodynamical simulations studying the environment in between galaxies \citep[e.g.,][]{Benitez-Llambay2013, PashaarXiv220404097P}
where large scale environmental elements such as sheets and filaments have been shown to strip gas in low mass halos. As of now this remains an exciting open question.

\section{Conclusions} \label{sec:conclusions}
 We investigated whether signs of two-halo conformity, where galaxies outside the dark matter halo of a primary galaxy share properties with it,  are present within 4 Mpc of the MW. This type of galactic conformity could help to explain the synchronized star formation found by \cite{Olsen2021} in the SFHs of 36 Local Volume dwarf galaxies from ANGST. 
To cover a contiguous volume up to 4 Mpc, we added Local Group galaxies for which there are publicly available SFHs, making  a cut on mass to include only Local Group galaxies more massive than the least massive LV dwarf in our sample.   This left us with a total sample of 53 dwarf galaxies with CMD-based SFHs. 

Having full SFHs for these galaxies granted us a unique opportunity to probe for conformity back in time by inferring galaxy properties inside our own lightcone. We selected lookback times of 0, 1, 2, and 4 Gyr. 
As seen in Figure~\ref{fig:sfr_sfr}, we found that at the time of observation, secondaries of high mass primaries showed more than 3$\sigma$ greater median SFR than the secondaries of low mass primaries at a distance of 2 to 3 Mpc from the primary galaxy. When looking at SFR of secondaries and splitting on primary sSFR, we saw with a nominal 2.8$\sigma$ significance that the median SFR of secondaries of low sSFR primaries is higher than their counterparts at the same distance. Given the overall statistical behavior of our conformity tests, we considered these to be probable detections of two-halo conformity.  

The quenched fraction of secondaries of primaries split on stellar mass and sSFR followed a similar pattern to that of SFR, but with nominal significances of 2.1$\sigma$ and 2.3$\sigma$ and occurring at 1 Gyr lookback time. In this case, when splitting on stellar mass of primaries, it is satellites of low stellar mass primaries at 1 Gyr that are more likely to be quenched than their counterparts in the 2 to 3 Mpc bin, and it is the secondaries of high sSFR primaries that are more likely to be quenched in this bin when splitting primaries on sSFR. This is consistent with our results examining SFR, as we do not expect secondaries of high sSFR/stellar mass primaries that have high SFR to be quenched. Given the overall statistical behavior of our conformity tests, we considered these to be possible detections of two-halo conformity within the Local Volume 1 Gyr ago.

  We have demonstrated a method of using the star formation histories of galaxies for tests of conformity that promises to reveal the co-evolution of galaxy properties through both space and time.  The analysis here has broad implications for better understanding galaxy evolution as influenced by external environment if applied to large surveys in future work where past conformity can now be explored by expanded tests reading galaxy properties from inside our light cone.

 We investigated the extent to which the Local Group could be contributing to conformity in the Local Volume by repeating the tests while removing Local Group galaxies. When we remove Local Group galaxies from our sample, we do not detect conformity in any time or separation bin for any property of the secondaries. Our results suggest that the evolution of galaxies in the nearby Local Volume has been complex.  It is also worth noting that none of our tests of one-halo conformity (via galaxy pairs at separations $<$1 Mpc) yielded a detection of conformity, although a few of the differences for SFR and quenched fraction approached the nominal 2$\sigma$ significance we required to declare a possible detection.  Galaxy formation simulations can be investigated to see how often our conformity signals could arise due to happenstance in small volumes rather than underlying physical correlations as well as whether we should have been able to detect one-halo conformity with the small number of $<$1 Mpc pairs available in this sample.    

Additionally, not seeing galactic conformity within the Local Volume sample that shows synchronized star formation illustrates that SFHs can reveal evolutionary trends within a population of galaxies that is not sensitive to pair properties. Just as using SFHs to reveal synchronized star formation in a population of galaxies provides insights into possible factors that can shape  galaxy growth over time, using SFHs in combination with tests of conformity allows us to probe these same factors by tracing the correlations of properties between pairs back in time. Using these two methods yields complementary information about evolution within a volume.

As the number and distance of detectable galaxies within the Local Volume grows with current and upcoming surveys such as JWST, Rubin, and Roman \citep[see ][]{Mao2021, Applebaum2021, Mutlu-Pakdil2021}, we can check if conformity signals appear on 2-3 Mpc scales in Local Volume sub-volumes not centered on the Local Group. Expanded tests of conformity over cosmic time as enabled by SFHs will allow us to leverage available observations to gain deep insights into galaxy evolution.

\section{Acknowledgements}

We would like to especially thank the reviewer for their insightful and helpful comments. We kindly thank Dan Weisz for his helpful correspondence. We further thank Anna Wright, Peter Kurczynski, Antara Basu-Zych, and Yao-Yuan Mao for their comments and useful discussion. CO and EG acknowledge support from NASA ADAP grant 80NSSC22K0487,HST grant HST-GO-15647.020-A, and
the U.S. Department of Energy, Office of Science, Office of High Energy Physics Cosmic Frontier Research program
under Award Number DE-SC0010008

\vspace{5mm}

\bibliography{library}{}

\begin{thebibliography}{}
\expandafter\ifx\csname natexlab\endcsname\relax\def\natexlab#1{#1}\fi
\providecommand{\url}[1]{\href{#1}{#1}}
\providecommand{\dodoi}[1]{doi:~\href{http://doi.org/#1}{\nolinkurl{#1}}}
\providecommand{\doeprint}[1]{\href{http://ascl.net/#1}{\nolinkurl{http://ascl.net/#1}}}
\providecommand{\doarXiv}[1]{\href{https://arxiv.org/abs/#1}{\nolinkurl{https://arxiv.org/abs/#1}}}

\bibitem[{{Applebaum} {et~al.}(2021){Applebaum}, {Brooks}, {Christensen},
  {Munshi}, {Quinn}, {Shen}, \& {Tremmel}}]{Applebaum2021}
{Applebaum}, E., {Brooks}, A.~M., {Christensen}, C.~R., {et~al.} 2021, \apj,
  906, 96, \dodoi{10.3847/1538-4357/abcafa}

\bibitem[{{Ayromlou} {et~al.}(2022){Ayromlou}, {Kauffmann}, {Anand}, \&
  {White}}]{Ayromlou2022}
{Ayromlou}, M., {Kauffmann}, G., {Anand}, A., \& {White}, S. D.~M. 2022, arXiv
  e-prints, arXiv:2207.02218.
\newblock \doarXiv{2207.02218}

\bibitem[{{Ben{\'\i}tez-Llambay} {et~al.}(2013){Ben{\'\i}tez-Llambay},
  {Navarro}, {Abadi}, {Gottl{\"o}ber}, {Yepes}, {Hoffman}, \&
  {Steinmetz}}]{Benitez-Llambay2013}
{Ben{\'\i}tez-Llambay}, A., {Navarro}, J.~F., {Abadi}, M.~G., {et~al.} 2013,
  \apjl, 763, L41, \dodoi{10.1088/2041-8205/763/2/L41}

\bibitem[{{Benson} {et~al.}(2002){Benson}, {Frenk}, {Lacey}, {Baugh}, \&
  {Cole}}]{Benson2002}
{Benson}, A.~J., {Frenk}, C.~S., {Lacey}, C.~G., {Baugh}, C.~M., \& {Cole}, S.
  2002, \mnras, 333, 177, \dodoi{10.1046/j.1365-8711.2002.05388.x}

\bibitem[{{Bray} {et~al.}(2016){Bray}, {Pillepich}, {Sales}, {Zhu}, {Genel},
  {Rodriguez-Gomez}, {Torrey}, {Nelson}, {Vogelsberger}, {Springel},
  {Eisenstein}, \& {Hernquist}}]{Bray2016}
{Bray}, A.~D., {Pillepich}, A., {Sales}, L.~V., {et~al.} 2016, \mnras, 455,
  185, \dodoi{10.1093/mnras/stv2316}

\bibitem[{{Brooks} {et~al.}(2013){Brooks}, {Kuhlen}, {Zolotov}, \&
  {Hooper}}]{Brooks2013}
{Brooks}, A.~M., {Kuhlen}, M., {Zolotov}, A., \& {Hooper}, D. 2013, \apj, 765,
  22, \dodoi{10.1088/0004-637X/765/1/22}

\bibitem[{{Campbell} {et~al.}(2015){Campbell}, {van den Bosch}, {Hearin},
  {Padmanabhan}, {Berlind}, {Mo}, {Tinker}, \& {Yang}}]{Campbell2015}
{Campbell}, D., {van den Bosch}, F.~C., {Hearin}, A., {et~al.} 2015, \mnras,
  452, 444, \dodoi{10.1093/mnras/stv1091}

\bibitem[{{Carlin} {et~al.}(2016){Carlin}, {Sand}, {Price}, {Willman},
  {Karunakaran}, {Spekkens}, {Bell}, {Brodie}, {Crnojevi{\'c}}, {Forbes},
  {Hargis}, {Kirby}, {Lupton}, {Peter}, {Romanowsky}, \&
  {Strader}}]{Carlin2016}
{Carlin}, J.~L., {Sand}, D.~J., {Price}, P., {et~al.} 2016, \apjl, 828, L5,
  \dodoi{10.3847/2041-8205/828/1/L5}

\bibitem[{{Carlin} {et~al.}(2021){Carlin}, {Mutlu-Pakdil}, {Crnojevi{\'c}},
  {Garling}, {Karunakaran}, {Peter}, {Tollerud}, {Forbes}, {Hargis}, {Lim},
  {Romanowsky}, {Sand}, {Spekkens}, \& {Strader}}]{Carlin2021}
{Carlin}, J.~L., {Mutlu-Pakdil}, B., {Crnojevi{\'c}}, D., {et~al.} 2021, \apj,
  909, 211, \dodoi{10.3847/1538-4357/abe040}

\bibitem[{{Carlsten} {et~al.}(2022){Carlsten}, {Greene}, {Beaton}, {Danieli},
  \& {Greco}}]{Carlsten2022}
{Carlsten}, S.~G., {Greene}, J.~E., {Beaton}, R.~L., {Danieli}, S., \& {Greco},
  J.~P. 2022, \apj, 933, 47, \dodoi{10.3847/1538-4357/ac6fd7}

\bibitem[{Dalcanton {et~al.}(2009)Dalcanton, Williams, Seth, Dolphin, Holtzman,
  Rosema, Skillman, Cole, Girardi, Gogarten, Karachentsev, Olsen, Weisz,
  Christensen, Freeman, Gilbert, Gallart, Harris, Hodge, de~Jong,
  Karachentseva, Mateo, Stetson, Tavarez, Zaritsky, Governato, \&
  Quinn}]{dalcanton}
Dalcanton, J.~J., Williams, B.~F., Seth, A.~C., {et~al.} 2009, The
  Astrophysical Journal Supplement Series, 183, 67

\bibitem[{{Deason} {et~al.}(2015){Deason}, {Wetzel}, {Garrison-Kimmel}, \&
  {Belokurov}}]{Deason2015}
{Deason}, A.~J., {Wetzel}, A.~R., {Garrison-Kimmel}, S., \& {Belokurov}, V.
  2015, \mnras, 453, 3568, \dodoi{10.1093/mnras/stv1939}

\bibitem[{Dolphin(2002)}]{dolphin}
Dolphin, A.~E. 2002, Monthly Notices of the Royal Astronomical Society, 332,
  91, \dodoi{10.1046/j.1365-8711.2002.05271.x}

\bibitem[{{Garrison-Kimmel} {et~al.}(2014){Garrison-Kimmel}, {Boylan-Kolchin},
  {Bullock}, \& {Lee}}]{Garrison-Kimmel2014}
{Garrison-Kimmel}, S., {Boylan-Kolchin}, M., {Bullock}, J.~S., \& {Lee}, K.
  2014, \mnras, 438, 2578, \dodoi{10.1093/mnras/stt2377}

\bibitem[{{Geha} {et~al.}(2017){Geha}, {Wechsler}, {Mao}, {Tollerud}, {Weiner},
  {Bernstein}, {Hoyle}, {Marchi}, {Marshall}, {Mu{\~n}oz}, \& {Lu}}]{Geha2017}
{Geha}, M., {Wechsler}, R.~H., {Mao}, Y.-Y., {et~al.} 2017, \apj, 847, 4,
  \dodoi{10.3847/1538-4357/aa8626}

\bibitem[{{Grand} {et~al.}(2021){Grand}, {Marinacci}, {Pakmor}, {Simpson},
  {Kelly}, {G{\'o}mez}, {Jenkins}, {Springel}, {Frenk}, \& {White}}]{Grand2021}
{Grand}, R. J.~J., {Marinacci}, F., {Pakmor}, R., {et~al.} 2021, \mnras, 507,
  4953, \dodoi{10.1093/mnras/stab2492}

\bibitem[{{Grogin} {et~al.}(2011){Grogin}, {Kocevski}, {Faber}, {Ferguson},
  {Koekemoer}, {Riess}, {Acquaviva}, {Alexander}, {Almaini}, {Ashby}, {Barden},
  {Bell}, {Bournaud}, {Brown}, {Caputi}, {Casertano}, {Cassata}, {Castellano},
  {Challis}, {Chary}, {Cheung}, {Cirasuolo}, {Conselice}, {Roshan Cooray},
  {Croton}, {Daddi}, {Dahlen}, {Dav{\'e}}, {de Mello}, {Dekel}, {Dickinson},
  {Dolch}, {Donley}, {Dunlop}, {Dutton}, {Elbaz}, {Fazio}, {Filippenko},
  {Finkelstein}, {Fontana}, {Gardner}, {Garnavich}, {Gawiser}, {Giavalisco},
  {Grazian}, {Guo}, {Hathi}, {H{\"a}ussler}, {Hopkins}, {Huang}, {Huang},
  {Jha}, {Kartaltepe}, {Kirshner}, {Koo}, {Lai}, {Lee}, {Li}, {Lotz}, {Lucas},
  {Madau}, {McCarthy}, {McGrath}, {McIntosh}, {McLure}, {Mobasher},
  {Moustakas}, {Mozena}, {Nandra}, {Newman}, {Niemi}, {Noeske}, {Papovich},
  {Pentericci}, {Pope}, {Primack}, {Rajan}, {Ravindranath}, {Reddy}, {Renzini},
  {Rix}, {Robaina}, {Rodney}, {Rosario}, {Rosati}, {Salimbeni}, {Scarlata},
  {Siana}, {Simard}, {Smidt}, {Somerville}, {Spinrad}, {Straughn}, {Strolger},
  {Telford}, {Teplitz}, {Trump}, {van der Wel}, {Villforth}, {Wechsler},
  {Weiner}, {Wiklind}, {Wild}, {Wilson}, {Wuyts}, {Yan}, \&
  {Yun}}]{Grogin2011CANDELS}
{Grogin}, N.~A., {Kocevski}, D.~D., {Faber}, S.~M., {et~al.} 2011, \apjs, 197,
  35, \dodoi{10.1088/0067-0049/197/2/35}

\bibitem[{{Hartley} {et~al.}(2015){Hartley}, {Conselice}, {Mortlock},
  {Foucaud}, \& {Simpson}}]{Hartley2015}
{Hartley}, W.~G., {Conselice}, C.~J., {Mortlock}, A., {Foucaud}, S., \&
  {Simpson}, C. 2015, \mnras, 451, 1613, \dodoi{10.1093/mnras/stv972}

\bibitem[{{Hearin} {et~al.}(2015){Hearin}, {Watson}, \& {van den
  Bosch}}]{HearinWatson2015}
{Hearin}, A.~P., {Watson}, D.~F., \& {van den Bosch}, F.~C. 2015, \mnras, 452,
  1958, \dodoi{10.1093/mnras/stv1358}

\bibitem[{Hopkins {et~al.}(2014)Hopkins, Kere{\v{s}}, O{\~{n}}orbe,
  Faucher-Gigu{\`{e}}re, Quataert, Murray, \& Bullock}]{Hopkins2014}
Hopkins, P.~F., Kere{\v{s}}, D., O{\~{n}}orbe, J., {et~al.} 2014, Monthly
  Notices of the Royal Astronomical Society, 445, 581,
  \dodoi{10.1093/mnras/stu1738}

\bibitem[{{Hopkins} {et~al.}(2018){Hopkins}, {Wetzel}, {Kere{\v{s}}},
  {Faucher-Gigu{\`e}re}, {Quataert}, {Boylan-Kolchin}, {Murray}, {Hayward},
  {Garrison-Kimmel}, {Hummels}, {Feldmann}, {Torrey}, {Ma},
  {Angl{\'e}s-Alc{\'a}zar}, {Su}, {Orr}, {Schmitz}, {Escala}, {Sanderson},
  {Grudi{\'c}}, {Hafen}, {Kim}, {Fitts}, {Bullock}, {Wheeler}, {Chan},
  {Elbert}, \& {Narayanan}}]{Hopkins2018}
{Hopkins}, P.~F., {Wetzel}, A., {Kere{\v{s}}}, D., {et~al.} 2018, \mnras, 480,
  800, \dodoi{10.1093/mnras/sty1690}

\bibitem[{{Ivezi{\'c}} {et~al.}(2019){Ivezi{\'c}}, {Kahn}, {Tyson}, {Abel},
  {Acosta}, {Allsman}, {Alonso}, {AlSayyad}, {Anderson}, {Andrew}, {Angel},
  {Angeli}, {Ansari}, {Antilogus}, {Araujo}, {Armstrong}, {Arndt}, {Astier},
  {Aubourg}, {Auza}, {Axelrod}, {Bard}, {Barr}, {Barrau}, {Bartlett}, {Bauer},
  {Bauman}, {Baumont}, {Bechtol}, {Bechtol}, {Becker}, {Becla}, {Beldica},
  {Bellavia}, {Bianco}, {Biswas}, {Blanc}, {Blazek}, {Blandford}, {Bloom},
  {Bogart}, {Bond}, {Booth}, {Borgland}, {Borne}, {Bosch}, {Boutigny},
  {Brackett}, {Bradshaw}, {Brandt}, {Brown}, {Bullock}, {Burchat}, {Burke},
  {Cagnoli}, {Calabrese}, {Callahan}, {Callen}, {Carlin}, {Carlson},
  {Chandrasekharan}, {Charles-Emerson}, {Chesley}, {Cheu}, {Chiang}, {Chiang},
  {Chirino}, {Chow}, {Ciardi}, {Claver}, {Cohen-Tanugi}, {Cockrum}, {Coles},
  {Connolly}, {Cook}, {Cooray}, {Covey}, {Cribbs}, {Cui}, {Cutri}, {Daly},
  {Daniel}, {Daruich}, {Daubard}, {Daues}, {Dawson}, {Delgado}, {Dellapenna},
  {de Peyster}, {de Val-Borro}, {Digel}, {Doherty}, {Dubois},
  {Dubois-Felsmann}, {Durech}, {Economou}, {Eifler}, {Eracleous}, {Emmons},
  {Fausti Neto}, {Ferguson}, {Figueroa}, {Fisher-Levine}, {Focke}, {Foss},
  {Frank}, {Freemon}, {Gangler}, {Gawiser}, {Geary}, {Gee}, {Geha}, {Gessner},
  {Gibson}, {Gilmore}, {Glanzman}, {Glick}, {Goldina}, {Goldstein}, {Goodenow},
  {Graham}, {Gressler}, {Gris}, {Guy}, {Guyonnet}, {Haller}, {Harris},
  {Hascall}, {Haupt}, {Hernandez}, {Herrmann}, {Hileman}, {Hoblitt}, {Hodgson},
  {Hogan}, {Howard}, {Huang}, {Huffer}, {Ingraham}, {Innes}, {Jacoby}, {Jain},
  {Jammes}, {Jee}, {Jenness}, {Jernigan}, {Jevremovi{\'c}}, {Johns}, {Johnson},
  {Johnson}, {Jones}, {Juramy-Gilles}, {Juri{\'c}}, {Kalirai}, {Kallivayalil},
  {Kalmbach}, {Kantor}, {Karst}, {Kasliwal}, {Kelly}, {Kessler}, {Kinnison},
  {Kirkby}, {Knox}, {Kotov}, {Krabbendam}, {Krughoff}, {Kub{\'a}nek},
  {Kuczewski}, {Kulkarni}, {Ku}, {Kurita}, {Lage}, {Lambert}, {Lange},
  {Langton}, {Le Guillou}, {Levine}, {Liang}, {Lim}, {Lintott}, {Long},
  {Lopez}, {Lotz}, {Lupton}, {Lust}, {MacArthur}, {Mahabal}, {Mandelbaum},
  {Markiewicz}, {Marsh}, {Marshall}, {Marshall}, {May}, {McKercher}, {McQueen},
  {Meyers}, {Migliore}, {Miller}, {Mills}, {Miraval}, {Moeyens}, {Moolekamp},
  {Monet}, {Moniez}, {Monkewitz}, {Montgomery}, {Morrison}, {Mueller},
  {Muller}, {Mu{\~n}oz Arancibia}, {Neill}, {Newbry}, {Nief}, {Nomerotski},
  {Nordby}, {O'Connor}, {Oliver}, {Olivier}, {Olsen}, {O'Mullane}, {Ortiz},
  {Osier}, {Owen}, {Pain}, {Palecek}, {Parejko}, {Parsons}, {Pease},
  {Peterson}, {Peterson}, {Petravick}, {Libby Petrick}, {Petry},
  {Pierfederici}, {Pietrowicz}, {Pike}, {Pinto}, {Plante}, {Plate}, {Plutchak},
  {Price}, {Prouza}, {Radeka}, {Rajagopal}, {Rasmussen}, {Regnault}, {Reil},
  {Reiss}, {Reuter}, {Ridgway}, {Riot}, {Ritz}, {Robinson}, {Roby}, {Roodman},
  {Rosing}, {Roucelle}, {Rumore}, {Russo}, {Saha}, {Sassolas}, {Schalk},
  {Schellart}, {Schindler}, {Schmidt}, {Schneider}, {Schneider}, {Schoening},
  {Schumacher}, {Schwamb}, {Sebag}, {Selvy}, {Sembroski}, {Seppala}, {Serio},
  {Serrano}, {Shaw}, {Shipsey}, {Sick}, {Silvestri}, {Slater}, {Smith},
  {Smith}, {Sobhani}, {Soldahl}, {Storrie-Lombardi}, {Stover}, {Strauss},
  {Street}, {Stubbs}, {Sullivan}, {Sweeney}, {Swinbank}, {Szalay}, {Takacs},
  {Tether}, {Thaler}, {Thayer}, {Thomas}, {Thornton}, {Thukral}, {Tice},
  {Trilling}, {Turri}, {Van Berg}, {Vanden Berk}, {Vetter}, {Virieux},
  {Vucina}, {Wahl}, {Walkowicz}, {Walsh}, {Walter}, {Wang}, {Wang}, {Warner},
  {Wiecha}, {Willman}, {Winters}, {Wittman}, {Wolff}, {Wood-Vasey}, {Wu},
  {Xin}, {Yoachim}, \& {Zhan}}]{Ivezic2019}
{Ivezi{\'c}}, {\v{Z}}., {Kahn}, S.~M., {Tyson}, J.~A., {et~al.} 2019, \apj,
  873, 111, \dodoi{10.3847/1538-4357/ab042c}

\bibitem[{{Iyer} {et~al.}(2019){Iyer}, {Gawiser}, {Faber}, {Ferguson},
  {Kartaltepe}, {Koekemoer}, {Pacifici}, \& {Somerville}}]{Iyer2019}
{Iyer}, K.~G., {Gawiser}, E., {Faber}, S.~M., {et~al.} 2019, \apj, 879, 116,
  \dodoi{10.3847/1538-4357/ab2052}

\bibitem[{{Jahn} {et~al.}(2022){Jahn}, {Sales}, {Wetzel}, {Samuel}, {El-Badry},
  {Boylan-Kolchin}, \& {Bullock}}]{Jahn2022}
{Jahn}, E.~D., {Sales}, L.~V., {Wetzel}, A., {et~al.} 2022, \mnras, 513, 2673,
  \dodoi{10.1093/mnras/stac811}

\bibitem[{Johnson {et~al.}(2013)Johnson, Weisz, Dalcanton, Johnson, Dale,
  Dolphin, de~Paz, Robert C.~Kennicutt, Lee, Skillman, Boquien, \&
  Williams}]{johnson}
Johnson, B.~D., Weisz, D.~R., Dalcanton, J.~J., {et~al.} 2013, The
  Astrophysical Journal, 772, 8.
\newblock \url{http://stacks.iop.org/0004-637X/772/i=1/a=8}

\bibitem[{{Kauffmann}(2015)}]{Kauffmann2015}
{Kauffmann}, G. 2015, \mnras, 454, 1840, \dodoi{10.1093/mnras/stv2113}

\bibitem[{{Kauffmann} {et~al.}(2013){Kauffmann}, {Li}, {Zhang}, \&
  {Weinmann}}]{Kauffmann2013}
{Kauffmann}, G., {Li}, C., {Zhang}, W., \& {Weinmann}, S. 2013, \mnras, 430,
  1447, \dodoi{10.1093/mnras/stt007}

\bibitem[{{Kawinwanichakij} {et~al.}(2016){Kawinwanichakij}, {Quadri},
  {Papovich}, {Kacprzak}, {Labb{\'e}}, {Spitler}, {Straatman}, {Tran}, {Allen},
  {Behroozi}, {Cowley}, {Dekel}, {Glazebrook}, {Hartley}, {Kelson}, {Koo},
  {Lee}, {Lu}, {Nanayakkara}, {Persson}, {Primack}, {Tilvi}, {Tomczak}, \& {van
  Dokkum}}]{Kawinwanichakij2016}
{Kawinwanichakij}, L., {Quadri}, R.~F., {Papovich}, C., {et~al.} 2016, \apj,
  817, 9, \dodoi{10.3847/0004-637X/817/1/9}

\bibitem[{{Knobel} {et~al.}(2015){Knobel}, {Lilly}, {Woo}, \&
  {Kova{\v{c}}}}]{Knobel2015}
{Knobel}, C., {Lilly}, S.~J., {Woo}, J., \& {Kova{\v{c}}}, K. 2015, \apj, 800,
  24, \dodoi{10.1088/0004-637X/800/1/24}

\bibitem[{{Koekemoer} {et~al.}(2011){Koekemoer}, {Faber}, {Ferguson}, {Grogin},
  {Kocevski}, {Koo}, {Lai}, {Lotz}, {Lucas}, {McGrath}, {Ogaz}, {Rajan},
  {Riess}, {Rodney}, {Strolger}, {Casertano}, {Castellano}, {Dahlen},
  {Dickinson}, {Dolch}, {Fontana}, {Giavalisco}, {Grazian}, {Guo}, {Hathi},
  {Huang}, {van der Wel}, {Yan}, {Acquaviva}, {Alexander}, {Almaini}, {Ashby},
  {Barden}, {Bell}, {Bournaud}, {Brown}, {Caputi}, {Cassata}, {Challis},
  {Chary}, {Cheung}, {Cirasuolo}, {Conselice}, {Roshan Cooray}, {Croton},
  {Daddi}, {Dav{\'e}}, {de Mello}, {de Ravel}, {Dekel}, {Donley}, {Dunlop},
  {Dutton}, {Elbaz}, {Fazio}, {Filippenko}, {Finkelstein}, {Frazer}, {Gardner},
  {Garnavich}, {Gawiser}, {Gruetzbauch}, {Hartley}, {H{\"a}ussler},
  {Herrington}, {Hopkins}, {Huang}, {Jha}, {Johnson}, {Kartaltepe},
  {Khostovan}, {Kirshner}, {Lani}, {Lee}, {Li}, {Madau}, {McCarthy},
  {McIntosh}, {McLure}, {McPartland}, {Mobasher}, {Moreira}, {Mortlock},
  {Moustakas}, {Mozena}, {Nandra}, {Newman}, {Nielsen}, {Niemi}, {Noeske},
  {Papovich}, {Pentericci}, {Pope}, {Primack}, {Ravindranath}, {Reddy},
  {Renzini}, {Rix}, {Robaina}, {Rosario}, {Rosati}, {Salimbeni}, {Scarlata},
  {Siana}, {Simard}, {Smidt}, {Snyder}, {Somerville}, {Spinrad}, {Straughn},
  {Telford}, {Teplitz}, {Trump}, {Vargas}, {Villforth}, {Wagner}, {Wandro},
  {Wechsler}, {Weiner}, {Wiklind}, {Wild}, {Wilson}, {Wuyts}, \&
  {Yun}}]{Koekemoer2011CANDELS}
{Koekemoer}, A.~M., {Faber}, S.~M., {Ferguson}, H.~C., {et~al.} 2011, \apjs,
  197, 36, \dodoi{10.1088/0067-0049/197/2/36}

\bibitem[{{Lacerna} {et~al.}(2018){Lacerna}, {Contreras}, {Gonz{\'a}lez},
  {Padilla}, \& {Gonzalez-Perez}}]{Lacerna2018}
{Lacerna}, I., {Contreras}, S., {Gonz{\'a}lez}, R.~E., {Padilla}, N., \&
  {Gonzalez-Perez}, V. 2018, \mnras, 475, 1177, \dodoi{10.1093/mnras/stx3253}

\bibitem[{{Lacerna} {et~al.}(2022){Lacerna}, {Rodriguez}, {Montero-Dorta},
  {O'Mill}, {Cora}, {Artale}, {Ruiz}, {Hough}, \&
  {Vega-Mart{\'\i}nez}}]{Lacerna2022}
{Lacerna}, I., {Rodriguez}, F., {Montero-Dorta}, A.~D., {et~al.} 2022, \mnras,
  513, 2271, \dodoi{10.1093/mnras/stac1020}

\bibitem[{{Mao} {et~al.}(2021){Mao}, {Geha}, {Wechsler}, {Weiner}, {Tollerud},
  {Nadler}, \& {Kallivayalil}}]{Mao2021}
{Mao}, Y.-Y., {Geha}, M., {Wechsler}, R.~H., {et~al.} 2021, \apj, 907, 85,
  \dodoi{10.3847/1538-4357/abce58}

\bibitem[{{McQuinn} {et~al.}(2010){McQuinn}, {Skillman}, {Cannon}, {Dalcanton},
  {Dolphin}, {Hidalgo-Rodr{\'\i}guez}, {Holtzman}, {Stark}, {Weisz}, \&
  {Williams}}]{McQuinn2010}
{McQuinn}, K. B.~W., {Skillman}, E.~D., {Cannon}, J.~M., {et~al.} 2010, \apj,
  724, 49, \dodoi{10.1088/0004-637X/724/1/49}

\bibitem[{{Mutlu-Pakdil} {et~al.}(2021){Mutlu-Pakdil}, {Sand}, {Crnojevi{\'c}},
  {Drlica-Wagner}, {Caldwell}, {Guhathakurta}, {Seth}, {Simon}, {Strader}, \&
  {Toloba}}]{Mutlu-Pakdil2021}
{Mutlu-Pakdil}, B., {Sand}, D.~J., {Crnojevi{\'c}}, D., {et~al.} 2021, \apj,
  918, 88, \dodoi{10.3847/1538-4357/ac0db8}

\bibitem[{{Olsen} {et~al.}(2021){Olsen}, {Gawiser}, {Iyer}, {McQuinn},
  {Johnson}, {Telford}, {Wright}, {Broussard}, \& {Kurczynski}}]{Olsen2021}
{Olsen}, C., {Gawiser}, E., {Iyer}, K., {et~al.} 2021, \apj, 913, 45,
  \dodoi{10.3847/1538-4357/abf3c2}

\bibitem[{{Pacifici} {et~al.}(2016){Pacifici}, {Oh}, {Oh}, {Lee}, \&
  {Yi}}]{Pacifici2016a}
{Pacifici}, C., {Oh}, S., {Oh}, K., {Lee}, J., \& {Yi}, S.~K. 2016, \apj, 824,
  45, \dodoi{10.3847/0004-637X/824/1/45}

\bibitem[{{Pasha} {et~al.}(2022){Pasha}, {Mandelker}, {van den Bosch},
  {Springel}, \& {van de Voort}}]{PashaarXiv220404097P}
{Pasha}, I., {Mandelker}, N., {van den Bosch}, F.~C., {Springel}, V., \& {van
  de Voort}, F. 2022, arXiv e-prints, arXiv:2204.04097.
\newblock \doarXiv{2204.04097}

\bibitem[{{Patel} {et~al.}(2018){Patel}, {Carlin}, {Tollerud}, {Collins}, \&
  {Dooley}}]{Patel2018}
{Patel}, E., {Carlin}, J.~L., {Tollerud}, E.~J., {Collins}, M. L.~M., \&
  {Dooley}, G.~A. 2018, \mnras, 480, 1883, \dodoi{10.1093/mnras/sty1946}

\bibitem[{{Phillips} {et~al.}(2014){Phillips}, {Wheeler}, {Boylan-Kolchin},
  {Bullock}, {Cooper}, \& {Tollerud}}]{Phillips2014}
{Phillips}, J.~I., {Wheeler}, C., {Boylan-Kolchin}, M., {et~al.} 2014, \mnras,
  437, 1930, \dodoi{10.1093/mnras/stt2023}

\bibitem[{{Putman} {et~al.}(2021){Putman}, {Zheng}, {Price-Whelan}, {Grcevich},
  {Johnson}, {Tollerud}, \& {Peek}}]{Putman2021}
{Putman}, M.~E., {Zheng}, Y., {Price-Whelan}, A.~M., {et~al.} 2021, \apj, 913,
  53, \dodoi{10.3847/1538-4357/abe391}

\bibitem[{{Samuel} {et~al.}(2022){Samuel}, {Wetzel}, {Santistevan}, {Tollerud},
  {Moreno}, {Boylan-Kolchin}, {Bailin}, \& {Pardasani}}]{Samuel2022}
{Samuel}, J., {Wetzel}, A., {Santistevan}, I., {et~al.} 2022, \mnras, 514,
  5276, \dodoi{10.1093/mnras/stac1706}

\bibitem[{Sin {et~al.}(2017)Sin, Lilly, \& Henriques}]{Sin2017}
Sin, L. P.~T., Lilly, S.~J., \& Henriques, B. M.~B. 2017, Monthly Notices of
  the Royal Astronomical Society, 471, 1192

\bibitem[{Tinker {et~al.}(2018)Tinker, Hahn, Mao, Wetzel, \&
  Conroy}]{Tinker2017}
Tinker, J.~L., Hahn, C., Mao, Y.-Y., Wetzel, A.~R., \& Conroy, C. 2018, Monthly
  Notices of the Royal Astronomical Society, 642

\bibitem[{{Tollerud} \& {Peek}(2018)}]{Tollerud2018}
{Tollerud}, E.~J., \& {Peek}, J.~E.~G. 2018, \apj, 857, 45,
  \dodoi{10.3847/1538-4357/aab3e4}

\bibitem[{{Wang} \& {White}(2012)}]{WangWhite2012}
{Wang}, W., \& {White}, S. D.~M. 2012, \mnras, 424, 2574,
  \dodoi{10.1111/j.1365-2966.2012.21256.x}

\bibitem[{{Weinmann} {et~al.}(2006){Weinmann}, {van den Bosch}, {Yang}, \&
  {Mo}}]{Weinmann2006}
{Weinmann}, S.~M., {van den Bosch}, F.~C., {Yang}, X., \& {Mo}, H.~J. 2006,
  \mnras, 366, 2, \dodoi{10.1111/j.1365-2966.2005.09865.x}

\bibitem[{Weisz {et~al.}(2014)Weisz, Dolphin, Skillman, Holtzman, Gilbert,
  Dalcanton, \& Williams}]{WeiszlgI2014}
Weisz, D.~R., Dolphin, A.~E., Skillman, E.~D., {et~al.} 2014, The Astrophysical
  Journal, 789, 147, \dodoi{10.1088/0004-637x/789/2/147}

\bibitem[{{Weisz} {et~al.}(2014){Weisz}, {Dolphin}, {Skillman}, {Holtzman},
  {Gilbert}, {Dalcanton}, \& {Williams}}]{Weisz2014reion}
{Weisz}, D.~R., {Dolphin}, A.~E., {Skillman}, E.~D., {et~al.} 2014, \apj, 789,
  148, \dodoi{10.1088/0004-637X/789/2/148}

\bibitem[{Weisz {et~al.}(2011{\natexlab{a}})Weisz, Dalcanton, Williams,
  Gilbert, Skillman, Seth, Dolphin, McQuinn, Gogarten, Holtzman, Rosema, Cole,
  Karachentsev, \& Zaritsky}]{weisz}
Weisz, D.~R., Dalcanton, J.~J., Williams, B.~F., {et~al.} 2011{\natexlab{a}},
  The Astrophysical Journal, 739, 5.
\newblock \url{http://stacks.iop.org/0004-637X/739/i=1/a=5}

\bibitem[{Weisz {et~al.}(2011{\natexlab{b}})Weisz, Dolphin, Dalcanton,
  Skillman, Holtzman, Williams, Gilbert, Seth, Cole, Gogarten, Rosema,
  Karachentsev, McQuinn, \& Zaritsky}]{weisz2}
Weisz, D.~R., Dolphin, A.~E., Dalcanton, J.~J., {et~al.} 2011{\natexlab{b}},
  The Astrophysical Journal, 743, 8.
\newblock \url{http://stacks.iop.org/0004-637X/743/i=1/a=8}

\bibitem[{{Wetzel} {et~al.}(2015){Wetzel}, {Tollerud}, \& {Weisz}}]{Wetzel2015}
{Wetzel}, A.~R., {Tollerud}, E.~J., \& {Weisz}, D.~R. 2015, \apjl, 808, L27,
  \dodoi{10.1088/2041-8205/808/1/L27}

\bibitem[{{York} {et~al.}(2000){York}, {Adelman}, {Anderson}, {Anderson},
  {Annis}, {Bahcall}, {Bakken}, {Barkhouser}, {Bastian}, {Berman}, {Boroski},
  {Bracker}, {Briegel}, {Briggs}, {Brinkmann}, {Brunner}, {Burles}, {Carey},
  {Carr}, {Castander}, {Chen}, {Colestock}, {Connolly}, {Crocker}, {Csabai},
  {Czarapata}, {Davis}, {Doi}, {Dombeck}, {Eisenstein}, {Ellman}, {Elms},
  {Evans}, {Fan}, {Federwitz}, {Fiscelli}, {Friedman}, {Frieman}, {Fukugita},
  {Gillespie}, {Gunn}, {Gurbani}, {de Haas}, {Haldeman}, {Harris}, {Hayes},
  {Heckman}, {Hennessy}, {Hindsley}, {Holm}, {Holmgren}, {Huang}, {Hull},
  {Husby}, {Ichikawa}, {Ichikawa}, {Ivezi{\'c}}, {Kent}, {Kim}, {Kinney},
  {Klaene}, {Kleinman}, {Kleinman}, {Knapp}, {Korienek}, {Kron}, {Kunszt},
  {Lamb}, {Lee}, {Leger}, {Limmongkol}, {Lindenmeyer}, {Long}, {Loomis},
  {Loveday}, {Lucinio}, {Lupton}, {MacKinnon}, {Mannery}, {Mantsch}, {Margon},
  {McGehee}, {McKay}, {Meiksin}, {Merelli}, {Monet}, {Munn}, {Narayanan},
  {Nash}, {Neilsen}, {Neswold}, {Newberg}, {Nichol}, {Nicinski}, {Nonino},
  {Okada}, {Okamura}, {Ostriker}, {Owen}, {Pauls}, {Peoples}, {Peterson},
  {Petravick}, {Pier}, {Pope}, {Pordes}, {Prosapio}, {Rechenmacher}, {Quinn},
  {Richards}, {Richmond}, {Rivetta}, {Rockosi}, {Ruthmansdorfer}, {Sandford},
  {Schlegel}, {Schneider}, {Sekiguchi}, {Sergey}, {Shimasaku}, {Siegmund},
  {Smee}, {Smith}, {Snedden}, {Stone}, {Stoughton}, {Strauss}, {Stubbs},
  {SubbaRao}, {Szalay}, {Szapudi}, {Szokoly}, {Thakar}, {Tremonti}, {Tucker},
  {Uomoto}, {Vanden Berk}, {Vogeley}, {Waddell}, {Wang}, {Watanabe},
  {Weinberg}, {Yanny}, {Yasuda}, \& {SDSS Collaboration}}]{SDSSYork}
{York}, D.~G., {Adelman}, J., {Anderson}, John~E., J., {et~al.} 2000, \aj, 120,
  1579, \dodoi{10.1086/301513}

\end{thebibliography}
\bibliographystyle{aasjournal}


\listofchanges

\end{document}